\documentclass[twoside]{article}
\usepackage{fleqn,espcrc2,epsfig}
\usepackage{graphicx}
\usepackage[figuresright]{rotating}

\newcommand{\ba}{\begin{eqnarray}}
\newcommand{\ea}{\end{eqnarray}}
\newcommand{\bas}{\begin{eqnarray*}}
\newcommand{\eas}{\end{eqnarray*}}
\newcommand{\be}{\begin{equation}}
\newcommand{\ee}{\end{equation}}
\newcommand{\bes}{\begin{equation*}}
\newcommand{\ees}{\end{equation*}}
\newcommand{\bi}{\begin{itemize}}
\newcommand{\ei}{\end{itemize}}

\newcommand{\bcentre}{\begin{center}}
\newcommand{\ecentre}{\end{center}}

\newcommand{\pme}[2]{{}^{+#1}_{-#2}}

\font\tenmsb=msbm10 scaled\magstep1
\font\sevenmsb=msbm7 scaled\magstep1
\font\fivemsb=msbm5 scaled\magstep1
\newfam\msbfam
\textfont\msbfam=\tenmsb
\scriptfont\msbfam=\sevenmsb
\scriptscriptfont\msbfam=\fivemsb

\newcommand{\Bhatt}{Bhattacharya \emph{ et al.} }

\newcommand{\order}[1]{{\mathcal O}(#1)}

\title{
\hfill\begin{minipage}{0pt}\scriptsize \begin{tabbing}
	\hspace*{\fill} Edinburgh-2001/13\\ \end{tabbing}\end{minipage}\\[8pt]
	\vspace{-3.0cm}
Heavy-light decay constants on the lattice}
\author{UKQCD Collaboration\\
	Presented by C.M.~Maynard\address{Department of Physics and Astronomy,
	University of Edinburgh EH9 3JZ, UK}} 
\begin{document}

\begin{abstract}
We present results for the leptonic decays of heavy-light mesons in
the quenched approximation and a preliminary result for $f_{D_s}$ with
dynamical fermions for $N_f=2$. We examine the systematic effects of
extrapolating the decay constants from the charm to the bottom quark
mass, and find this extrapolation to be under control. We present a
preliminary result for $f_{D_s}^{N_f=2}$ and compare to
$f_{D_s}^{N_f=0}$ matched to have the same lattice spacing set by the
Sommer scale, $r_0$. We find no significant difference.

\end{abstract}

\maketitle

\section{INTRODUCTION}
The decay constants of the B and D mesons are important in
phenomenology. The combination $f_B\sqrt{ B_B}$, $f_{B_s}\sqrt{ B_{B_s}}$, 
and their ratio, play a crucial role in extracting CKM
matrix elements from experimental data. The B-factory experiments
BaBar and Belle will measure $B^0-\bar{B^0}$ mixing quantities, but 
measuring $f_B$ will be very difficult.

On the lattice, $f_B$ is a mature calculation, and less extensive than
$B_B$, so we can test the systematics of a heavy-light lattice
calculation.  The ``world average'' for $f_B$~\cite{Bernard_2000} in
the quenched approximation is $f_B=175(20)$ MeV and with dynamical
fermions (DF), $f_B^{N_f=2}=200(30)$ MeV (see
also~\cite{Ryan_2001}). This suggests a strong dependence on sea quark
effects. However, it is not clear how to disentangle different
systematic effects.

\section{QUENCHED DECAY CONSTANTS}
These results have been partially reported in
\cite{np_imp_fb,chrism_2000}.  We use the non-perturbatively (NP)
$\order{a}$ improved Wilson quark action at two values of the
coupling, $\beta=6.2$ and $\beta=6.0$. The light quarks have masses
around strange, and the heavy quarks around charm. In the NP
improvement scheme, the axial current is given by
\be
\label{eqn:axial_current}
  A_\mu^{\rm R} =  Z_A(1+{b_A} am_q)( A_\mu + a {c_A} \partial_\mu P_s)
\ee
We use non-perturbative determinations of the coefficients $Z_A$,
$b_A$ and $c_A$ by \Bhatt~\cite{Bhatta_2000}. In \cite{np_imp_fb}
we show that the value of the decay constant depends strongly on the
value of the mixing coefficient $c_A$. These coefficients are only
defined up to $\order{a}$, e.g. compare
the ALPHA~\cite{alpha_np2} and \Bhatt values of $c_A$ at $\beta=6.0$.
$c_A^{\rm ALPHA}=-0.083(5)$ and $c_A^{\rm Bhatt}=-0.037(4)(7)$. However,
observables computed with these coefficients have the same continuum limit,
provided each set of coefficients is determined consistently.

We use continuum Heavy Quark Symmetry (HQS) to motivate the form of
the extrapolation to the $b$ quark mass;
\be
\label{eqn:HQS}
  \Phi(M)=\Theta(M) f \sqrt{M}=\gamma\left ( 1 + \frac{\delta}{M} + \frac{\eta}{M^2} \right)
\ee
where $M$ is the pseudoscalar meson mass, $f$ the pseudoscalar decay constant
defined by
\be
  \langle 0 | A_\mu | P_s(\vec{p})\rangle = ifp_\mu
\ee
and $\Theta$ is the logarithmic correction from QCD.  

\begin{figure}[!ht]
\vspace{-0.5cm}
\begin{center}
\epsfig{file=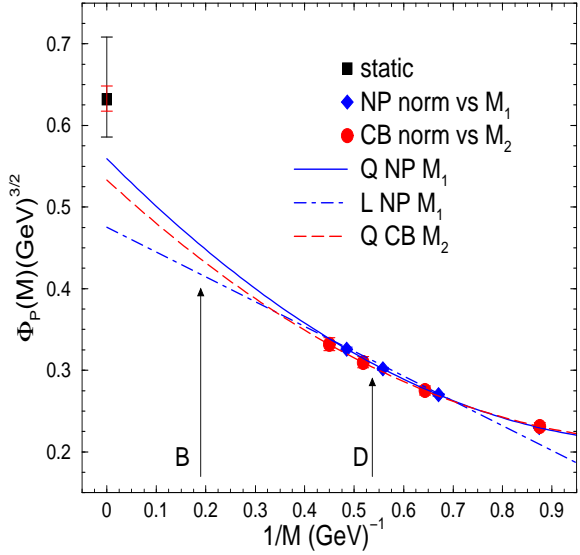,height=7.5cm,width=7.5cm}
\vspace{-1.5cm}
\caption{HQS extrapolation of $\Phi$ at $\beta=6.0$}
\label{fig:static}
\end{center}
\vspace{-1.0cm}
\end{figure}

This extrapolation is shown in figure~\ref{fig:static}.  We have also
computed the value of $\Phi$ in the static limit, but have not
included the improvement term. We note that $c_A$ is negative, which
will reduce the value of $\Phi$. We do not interpolate from charm, but
use the static point as a check in the extrapolation.

Also shown in figure~\ref{fig:static} is a change of normalisation of
the axial current labelled CB, as suggested by
Bernard~\cite{Bernard_2000},
\be
   (A_0^R)_\mathrm{CB}=
	Z_A  \left (1 + b_A am_q + 2 a c_A {\mathcal R} \right)^{1/2}A_0
\ee
where ${\mathcal R}=\partial_0 \langle 0 | P | P \rangle/m_q\langle 0 | A_0 | P \rangle $
This differs from the normalisation in equation~(\ref{eqn:axial_current}) at
$\order{a^2}$, but has a finite static limit. This is plotted against $M_2$,
the kinetic mass, 
$M_2^{-1}=\left.\partial^2 E/\partial p_i^2 \right|_{\vec{p}=0}$
%\be
%  M_2^{-1}=\left.\partial^2 E/\partial p_i^2 \right|_{\vec{p}=0}
%\ee
There is a small difference between our standard extrapolation and that
labelled CB. 

\begin{figure}[!ht]
\vspace{-0.75cm}
\begin{center}
\epsfig{file=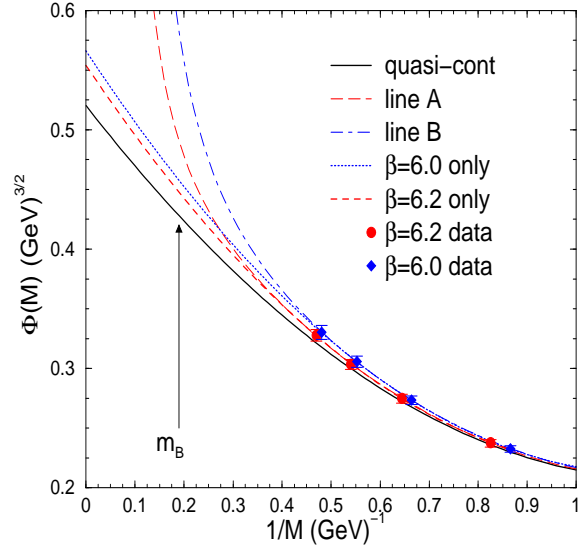,height=7.5cm,width=7.5cm}
\vspace{-1.5cm}
\caption{Quasi-Continuum extrapolation of $\Phi$}
\label{fig:quasi-cont}
\end{center}
\vspace{-1.0cm}
\end{figure}

To estimate the effect of mass dependent lattice artefacts directly
from the data, we add terms to equation (\ref{eqn:HQS}) that model
the lattice artefacts,
\be
\Phi^\prime(M,a)=\Phi(M) + \gamma\left( \varepsilon (aM)^2 + \zeta (aM)^3
				 \right)
\ee
We then fit to both lattice spacings simultaneously. Shown in
figure~\ref{fig:quasi-cont} is the data at both values of $\beta$.
The fit parameters $\varepsilon$ and $\zeta$ are small and consistent
with zero. Line A(B) shows the function $\Phi^{\prime}(M,a)$ with $a$ at
$\beta=6.2(6.0)$. These show that lattice artefacts indeed
do blow up if extrapolated.  However, the lines labelled 
``$\beta=6.2(6.0)$ only'' and 
``quasi-continuum'' (QC) ($\Phi^\prime(M,0)$), lie close
together. 
Moreover $f_B^{\rm QC}=186(12)$ MeV, is close to 
the values of $f_B$ at fixed lattice spacing, albeit with larger errors,
which implies mass dependent lattice artefacts are not spoiling the
extrapolation.

We estimate the systematic uncertainties from the quantity used
to fix $a$, ($r_0, m_\rho$), linear vs quadratic HQS
extrapolation, quasi-continuum extrapolation and the difference
between different lattice spacings. We observe
good scaling between different lattice spacings. The values for the
quenched decay constants are as follows, with the scale set by
$f_\pi$. The errors are statistical and systematic;
\begin{table}[!ht]
\vspace{-0.75cm}
\begin{center}
\begin{tabular}{r@{$=$}lr@{$=$}l}
 $ f_B$&$195(6)^{+24}_{-23}\ {\rm MeV} $&
 $ f_D$&$206(4)^{+17}_{-10}\ \rm{MeV}$\\[1ex]
 $ f_{B_S}$&$220(6)^{+23}_{-28}\ \mathrm{MeV}$ &
 $ f_{D_S}$&$229(3)^{+23}_{-12}\ \mathrm{MeV} $ \\ [1ex]
 $ \frac{f_{B_s}}{f_B} $&$ 1.13(1)\pme{1}{5} $ &  
 $ \frac{f_{D_s}}{f_D} $&$ 1.11(1)\pme{1}{3} $  \\ [1ex]
 $ f_K$&$150(3)^{+12}_{-\ 8}\  \rm{MeV}$ & \multicolumn{2}{c}{  }\\
\end{tabular}
\end{center}
\vspace{-1.25cm}
\end{table}

\section{DECAY CONSTANTS WITH $N_f=2$}
The UKQCD collaboration has a set of matched ensembles \cite{Action_tuning},
where the lattice spacing (as measured by $r_0$) is fixed to be the same,
as $\beta$ and $m_q^{\rm sea}$ are varied, including a quenched set. 
These ensembles are NP $\order{a}$ improved.
Measuring the decay constants on these ensembles will enable us to study
the effect of unquenching whilst keeping everything else fixed.
Furthermore, we intend to study the systematics of different heavy quark
methods by comparing NRQCD, FNAL formalism and charm plus extrapolation 
directly on the same data sets. In particular it is interesting to see
if the ratio $f_B^{N_f=2}/f_B^{N_f=0}$ is the same for each method.

We present a preliminary result for $f_{D_s}$ at fixed lattice
spacing, comparing the $N_f=2$ value with the quenched value.  The
simulation details are shown in Table~\ref{tab:sim_details}.

\begin{table}[!ht]
\vspace{-0.8cm}
\begin{center}
\caption{Simulation details}
\label{tab:sim_details}
\begin{tabular}{ccc}
\hline
& $N_f=2$ & $N_f=0$ \\\hline
$\beta$, $\kappa^{\rm sea}$& $5.2$, $0.1350$ & $5.93$, $0$ \\
$m_\pi/m_\rho$& $0.70(1)$ & $1$ \\
cfgs & $ 202$ & $ 117$\\
$a^{-1}$, volume & \multicolumn{2}{c}{$0.1$ fm, $16^3\times 32$} \\
valence light & \multicolumn{2}{c}{3 $\sim$ strange}\\
valence  heavy & \multicolumn{2}{c}{4 $\sim$ charm}\\
$c_A$ BPT& $-0.0164$ & $-0.0132$ \\
$c_A$ NP& $-0.04(2)$ & $-0.032(14)$ \\\hline
\end{tabular}
\end{center}
\vspace{-1.25cm}
\end{table}

\begin{figure}[!ht]
\vspace{-0.5cm}
\begin{center}
\epsfig{file=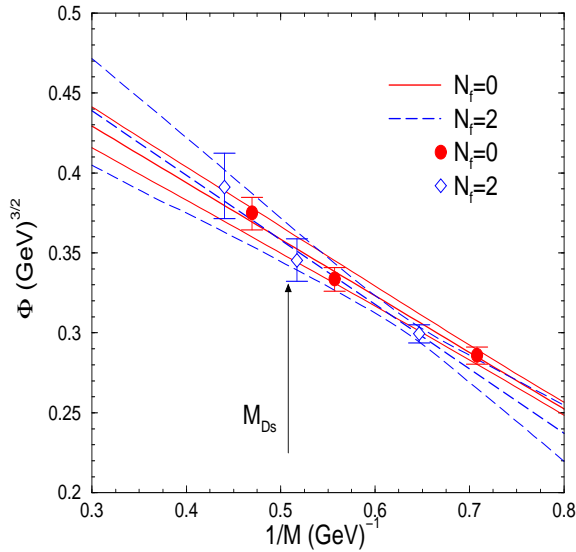,height=7.5cm,width=7.5cm}
\vspace{-1.5cm}
\caption{Comparison between DF and quenched data}
\label{fig:DF}
\end{center}
\vspace{-1.0cm}
\end{figure}

The values of the improvement coefficients have been determined to one-loop
in perturbation theory~\cite{alpha_np4_bA}. At one-loop these are independent
of sea quark effects. We evaluate these in ``boosted'' perturbation theory.
The interpolation to charm is shown in figure~\ref{fig:DF}. We see no
significant difference.

There is a preliminary NP
determination~\cite{collins_cA_2001}. Table~\ref{tab:fDs} shows the
dependence of $f_{D_s}$ on the value of $c_A$.  The entry $P_S$ shows
the value of the pseudoscalar density contribution.
\begin{table}[!ht]
%\vspace{-1cm}
\begin{center}
\caption{$f_{D_s}$ in MeV and its dependence on $c_A$.}
\label{tab:fDs}
\begin{tabular}{lcc}
\hline$c_A$& $N_f=2$ & $N_f=0$ \\\hline
BPT & $253(9)$ & $253(6)$ \\
NP  & $237(9)$ & $241(6)$ \\
0 & $264(9)$ & $262(6)$ \\
$P_S$ &$693(24)$ & $629(14)$ \\\hline
\end{tabular}
\end{center}
\vspace{-1cm}
\end{table}

Other than a small difference in the improvement term there is no difference.
One reason may be the sea quarks we have used are still rather heavy. 
The Sommer scale $r_0$ is a short distance quantity and the decay constant,
being proportional to the wavefunction at the origin is also a short distance
quantity. We have fixed $a$ with a short distance
quantity, thus the short distance physics will be similar for both
quenched and unquenched. It is not surprising that we observe no
difference. The MILC collaboration \cite{Bernard_2001} see a similar
pattern. Little difference when the scale is set by $r_1$ (related to $r_0$)
and significant $m_q^{\rm sea}$ dependence when the scale is set by $m_\rho$.
The decay constants are dimensionfull quantities, so they are
are sensitive to the scale setting ambiguity in the quenched approximation.
The change in fDs from unquenching depends very strongly
on how we choose to set the scale.

We acknowledge the following grants; EPSRC GR/K41663,
PPARC PPA/G/S/1999/00022 and PPA/GS/1997/00655
and EU HPRN-CT-2000-00145 Hadrons/LatticeQCD.
CMM acknowleges PPA/P/S/1998/00255 and thanks C.~Bernard, P.~Boyle,
S.~Collins, J.Hein, C.McNeile, and S.~Ryan for discussions and advice.

\end{document}